\documentclass[nofootinbib,aps,twocolumn,showpacs,amsmath,amssymb,unsortedaddress,a4paper]{revtex4}
\usepackage{amssymb}
\usepackage{amsmath}
\usepackage{epsfig}
\usepackage{amsfonts}
\usepackage{color}
\usepackage{float}
\usepackage{latexsym}
\usepackage{mathrsfs}
\usepackage{feynmf}
\usepackage{balance}
\usepackage{multirow}
\begin{document}
\title{ The Cabello Nonlocal Argument is Stronger Control than the Hardy Nonlocal Argument for Detecting Post-Quantum Correlations}
\author{Ali Ahanj}
\email{a.ahanj@khayyam.ac.ir; ahanj@ipm.ir}
\affiliation{Department of Physics, Khayyam University, Mashhad, Iran.}
\affiliation{Foundations of Physics Group, School of Physics, Institute for Research in Fundamental Sciences (IPM), P. O. Box 19395-5531, Tehran, Iran.}
%%%%%%%%%%%%%%%%%%%%%%%%%%%%%%%%%%%%%%%%%%%%%%%%%%%%%%%%%%%%%%%%%%%%%%%%%%%%%%%%%%%%%%%%
\begin{abstract}
In this paper, we study the Hardy nonlocal argument (HNA) and the Cabello nonlocal argument (CNA) under the Information Causality (IC), Macroscopic Locality (ML) and Local Orthogonality (LO) principles in the context of Local Randomness. We see that, in the context of all the possibilities of local randomness, the gap between the quantum mechanics and the above principles, in the CNA is larger than the HNA. Therefore the CNA is stronger control than the HNA for detecting post-quantum nosignalling correlations.
\end{abstract}
%%%%%%%%%%%%%%%%%%%%%%%%%%%%%%%%%%%%%%%%%%%%%%%%%%%%%%%%%%%%%%%%%%%%%%%%%%%%%%%%%%%%
\maketitle
 \section{Introduction}
 It is known that the value of Bell-CHSH \cite{bell,chsh} quantity, exceeding the local hidden variable theorems (LHVT) bound, quantifies a measure of nonlocality. Therefore, measurement of nonlocality within the quantum mechanics (QM) is limited by the Cirel'son bound $2\sqrt{2}$ \cite{tsi80} and the bound of nonlocality imposed by the no-signaling (NS) principle (PR box) is 4 \cite{pr}. On the other hand, Hardy \cite{har92,har93} introduced a non-locality theorem to prove Bell's theorem without inequality. The Hardy's argument of ``nonlocality without inequality"  has been considered to be the `` best version of Bell's theorem" \cite{mermin}. After that the authors \cite{cab02,lin,choud} shown that the Hardy nonlocality argument (HNA) is a special case of the Cabello nonlocality argument (CNA). One important difference between HNA and CNA is that a mixed two-qubit entangled state can never exhibit HNA, but they can exhibit CNA\cite{kar}. We know that the maximum success probability for the HNA and the CNA in LHVTs are zero, but in the QM, the corresponding values are $0.09$ \cite{har92,har93,rabel} and $0.11$\cite{kunkri} respectively. On the other hand, it has been shown that the upper bound of success probability of HNA and CNA in the general NS theorem is $0.5$ \cite{choud}.\\
 The natural question arise is that; why is QM not more nonlocal than it is$?$  Why quantum not violate Bell's inequality(BI) and HNA and CNA more than the number $ 2 \sqrt{2}$, 0.09 and 0.11 respectively?\\
  In the recent years, several attempts have been taken to get an answer to this fundamental question by using some physical or information theoretical principles. Some of these principles are non trivial communication complexity \cite{van,brassard}, Information Causality (IC) \cite{nature}, Macroscopic Locality (ML) \cite{nav09}, Exclusivity principle \cite{exp}, Relativistic Causality in the Classical limit \cite{roh1,roh2}, the uncertainty principle along with steering \cite{unp} and complementarity principle \cite{comp}. It is interesting, some of these principles like IC and ML can explain the Cirel'son bound of Bell-CHSH, but they can not explain the limit HNA and CNA in the quantum theory \cite{ahanj, lohardy}. It has been shown that under applying the IC, the upper bound on the success probability of both HNA and CNA is almost $0.2071$ \cite{ahanj}, and under applying the ML condition they are $0.2062$  \cite{lohardy}.\\
  Very soon, it has been shown that for a more than two subsystem correlations, no bipartite physical principle (like IC and ML) can completely recognize the full set of QM correlations \cite{gallego,das13}. In order the solve this problem, some new principles like Local Orthogonality (LO) is introduced by Fritz and et.al{\cite{fritz}}. Next, the authors in ref \cite{lohardy} found that the maximum success probability of HNA under the LO principle is $0.177$ which is closer to the QM value, but in the case CNA, it is same as that obtained from IC principle (0.207).\\ On the other hand, Gazi and et.al \cite{gazi} shown that in terms of local randomness (LR) of the observable for the Hardy's correlations, there is also a gap between QM and IC condition and this  gap in the context of CNA is much larger than the HNA case \cite{golnaz}. In this paper, we extend this approach to  ML and LO principles and get some interesting results.\\ This article is organized as follows. In Sec.II, we overview the HNA and the CNA  under the no-signaling condition. In Sec.III and IV, we briefly review the IC, ML and LO principles and  obtain the maximum success probability HNA and CNA under these principles in the context of LR, and  we bring our conclusions in sec V. The all inequalities are sufficient for obtaining the upper bound of the HNA and CNA under LO principle are presented in the appendices A and B. Finally the details of the MATLAB programs are shown in the appendix C.
\section{HNA and CNA  under the NS condition}
 Consider the convex set of bipartite no-signaling correlations with binary inputs and binary outputs for each party in a $2^4$ dimensional vector space. Let $P_{ab|xy}$ denote the joint probabilities, where $ a, b\in\{0,1\}$ and $x, y\in\{0,1\}$ denote outputs and inputs of two parties (we call them Alice and Bob)  respectively.
 The joint probabilities satisfy the positivity constraints:
\begin{equation}\label{positivity}
  0 \leq P_{ab|xy}\leq 1 ~~~\forall a,b,x,y \in \{0,1\}
\end{equation}
and also satisfy the normalization  constraints:
\begin{equation}\label{normalization}
 \sum_{a,b=0}^{1} P_{ab|xy}=1 ~~~\forall x,y \in \{0,1\} .
 \end{equation}
 On the other hand, since the parties ( Alice and Bob) are spatially separate, causality and relativity imply that the joint probabilities satisfy the no-signaling (NS) correlations. The NS  conditions imposing that the choice of measurement by Alice (or Bob) can not affect the outcome distributions of Bob ( or Alice). In the other words, the marginal conditional probabilities $p_{a|x}$ and $p_{b|y}$ must be independent of $y$ and $x$, respectively. The NS constraints can be express:
\begin {eqnarray}
\sum_{b} P_{ab|x0}&=&\sum_{b} P_{ab|x1}=p_{a|x} ~~\forall a, x\in \{0,1\}\label{ns1}\\
\sum_{a} P_{ab|0y}&=&\sum_{a} P_{ab|1y}=P_{b|y} ~~\forall b, y\in\{0,1\}\label{ns2}.
\end{eqnarray}
  The full set of normalization and  the NS correlations form an eight dimensional polytope structure \cite{bar05} that we call it, NS polytope.  We can represent these types of correlations by a $4\times4$ correlation matrix with 8 independent parameters:
  \begin{equation}\label{matrix}
\left(
  \begin{array}{cccc}
   e_1 & f_1-e_1 & g_1-e_1 & 1+e_1-f_1-g_1 \\
   e_2 & f_1-e_2 & g_2-e_2 & 1+e_2-f_1-g_2 \\
   e_3 & f_2-e_3 & g_1-e_3 & 1+e_3-f_2-g_1 \\
   e_4 & f_2-e_4 & g_2-e_4 & 1+e_4-f_2-g_2 \\
 \end{array}
\right)
\end{equation}
where the parameters $0 \leq e_k\leq 1$, $0 \leq f_i \leq 1$ and $0\leq g_i \leq 1$ ($k=1,2,3,4$ and $i=1,2 $) make the elements of matrix and the positivity constraint  is guaranteed by the condition: $$  max\{0, g+f-1\} \leq e \leq  min\{f,g\}.$$ The sets of LHVT and QM correlations are strictly contained within the NS polytope, but notice that the set of LHVT correlations like NS correlation forms a convex polytope, whereas the set of quantum correlations is convex but is not a polytope \cite{tsi80, jonathan}. The eight dimensional NS polytope has 24 vertices contains 16 local vertices:
\begin{equation}\label{local}
  P^{\alpha\beta\gamma\delta}_{ab|xy}=\delta_{(a=\alpha x \oplus \beta)}~\delta_{(b=\gamma y \oplus \delta)}
\end{equation}
and the eight nonlocal vertices:
\begin{equation}\label{nonlocal}
  P^{\alpha\beta\gamma}_{ab|xy}=\frac{1}{2}\delta_{a\oplus b= xy\oplus \alpha x \oplus \beta y \oplus \gamma}
\end{equation}
where $\alpha,\beta,\gamma,\delta\in \{0,1\}$ and $\oplus$ is addition modulo $2$. Now, we consider joint probabilities satisfying the Hardy-Cabello argument:
\begin{eqnarray}
  P_{ab|xy} &=& q_{1}, \nonumber\\
  P_{a^{\prime} \bar{b}|\bar{x} y} &=& 0, \nonumber\\
  P_{\bar{a}b^{\prime}|x\bar{y}} &=& 0, \nonumber\\
  P_{a^{\prime} b^{\prime}|\bar{x}\bar{y}} &=& q_{4},
\end{eqnarray}
where $\bar{\alpha}$ denotes complement of $\alpha$ (~$\bar{\alpha}=1\oplus \alpha)$ .  These equations form the basis of the Hardy-Cabello nonlocality argument. It can easily be shown that these equations contradict local realism if $q_1 < q_4$. Whenever $q_1=0$, the Cabello's argument reduces to the Hardy's argument. In the remaining part of this paper, without loss of generality we consider the following form of the Hardy-Cabello correlation:
$P_{01|01}=q_{1}>0$, $P_{00|11}=q_{2}=0$,  $P_{10|00}=q_{3}=0$, $P_{00|10}=q_{1}$. If we consider $q_1=0$ (Hardy's argument), then the above conditions, can be written as a convex combination of the $ 5 $ of the $16$ local vertices and one of the 8 nonlocal vertices \cite{ahanj, gazi, golnaz}:
\begin{eqnarray}\label{hardy}
  P^{H}_{ab|xy}&=&c_1 P^{0101}_{ab|xy}+c_2 P^{0111}_{ab|xy}+c_3 P^{1000}_{ab|xy}+c_4 P^{1011}_{ab|xy}\nonumber\\&+&c_5 P^{1101}_{ab|xy}+c_6 P^{000}_{ab|xy}
\end{eqnarray}
where $\sum^{6}_{i=1} c_i=1$. By placing the local matrices $P^{0101}_{ab|xy},P^{0111}_{ab|xy}, P^{1000}_{ab|xy}, P^{1011}_{ab|xy} , P^{1101}_{ab|xy}$  using  Eq (\ref{local}) and  the nonlocal matrix $P^{000}_{ab|xy}$  from Eq (\ref{nonlocal}) in the above equation, We get a  $4\times4$ Hardy correlation nonlocal matrix:
\begin{equation}\label{hardy matrix}
  \left(
    \begin{array}{llll}
      c_3& c_4 & 0&c_{1,2,5}\\
      c_{3,4}& 0 & c_2& c_{1,5}\\
      0 & c_5 & c_3& c_{1,2,4}\\
      0&c_5 & c_{2,3,4}& c_{1} \\
    \end{array}
  \right)
%\end{equation*}
%\begin{equation}\label{hardy matrix}
  +\frac{1}{2}\left(
    \begin{array}{llll}
      c_{6} &0 &0&c_{6}\\
      c_6& 0 & 0& c_6\\
      c_6 & 0 &0& c_6\\
      0&c_{6}&c_{6}&0 \\
    \end{array}
  \right)
\end{equation}
where $c_{i,j,k}\equiv c_i+c_j+c_k$. The Cabello's non-locality argument ($q_1\neq 0$),  can be written as a convex combination of the above $ 6 $ vertices which satisfies Hardy's conditions along with another four local vertices $P^{0001}_{ab|xy}$,~$P^{0010}_{ab|xy}$,~$P^{1001}_{ab|xy}$,~$P^{1010}_{ab|xy}$ and one nonlocal vertex $P^{110}_{ab|xy}$ :
 \begin{eqnarray}\label{cabelo}
  P^{C}_{ab|xy}\nonumber&=&P^{H}_{ab|xy}+c_7 P^{0001}_{ab|xy}+c_8 P^{0010}_{ab|xy}\\&+&c_9 P^{1001}_{ab|xy}+c_{10} P^{1010}_{ab|xy}+c_{11} P^{110}_{ab|xy}
\end{eqnarray}
where the expression $P^{H}_{ab|xy}$ is given in Eq( \ref{hardy}) and coefficients $c_i$'s satisfy $\sum_{i=1}^{11}c_i=1$. Also by using Eqs (\ref{local},\ref{nonlocal}) the correlation matrix for the Cabello's non-signaling boxes can be written as \cite{ahanj, gazi}:
\begin{equation*}
  \left(
    \begin{array}{llll}
      c_{3,8,10} & c_{4,7,9} & 0&c_{1,2,5}\\
      c_{3,4} & c_{7,8,9,10} & c_2& c_{1,5}\\
      c_8 & c_{5,7} & c_{3,10}& c_{1,2,4,9}\\
      0 &c_{5,7,8}&c_{2,3,4}&c_{1,9,10} \\
    \end{array}
  \right)
\end{equation*}
\begin{equation}\label{cabelo matrix}
  +\frac{1}{2}\left(
    \begin{array}{llll}
     c_{6,11} &0 &0&c_{6,11}\\
      c_6 & c_{11} & c_{11}& c_6\\
      c_6& c_{11} &c_{11}& c_6\\
      0&c_{6,11}&c_{6,11}&0. \\
    \end{array}
  \right)
\end{equation}\\
It has worth, notice here that while the upper bound of the success probability of the HNA and the CNA in QM is $0.09$ \cite{har92,har93} and $0.11$ \cite{kunkri} respectively, but under imposing the NS condition \cite{choud,jose}, by substituting $c_i=\delta_{i,6}$, they are same and equal to 0.5 (please see the TABLE I). Now, in the next section, we briefly review the IC, ML and LO principles.
\section{ Hardy's and Cabello's non-locality under IC, ML and LO principles}
\subsection{Information Causality Principle}
Pawlowaski and et.al \cite{nature} introduced Information Causality (IC) principle for underestanding the quantum mechanical bound on nonlocal correlations. Alice and Bob, who are separated in space, have access to non-signaling resources such as shared randomness, entanglement or PR boxes. Alice receives a randomly generated N-bit string $\vec{x}=(x_0,x_1,...,x_{N-1})$ while Bob is asked to guess Alice's $i$-th bit where $i$ is randomly chosen from the set $\vec{y}=\{0,1,2,...,N-1\}$. Alice then sends $M$ classical bits to Bob $(M<N)$ and let Bob's answer  be denoted by $\beta_i$. Then, the amount of the information about the variable $x_i$ of Alice's input potentially gained by Bob is measured by:
 \begin{equation}
 I \equiv \sum I(x_i:\beta|y=i) \geq N-\sum_{k=1}^{N} h(p_k)
 \end{equation}
 where $I(x_i:\beta|y=i)$ is Shanon mutual information between $x_i$ and $\beta$, and $p_k$ is the probability that $x_i=\beta$ and inequality can be proved by Fano inequality. The statement of the IC is that the total potential information \cite{nature} about Alice's bit string $\vec{x}$ accessible to Bob cannot exceed the volume of message  received from Alice, i.e., $ I= \sum_{i=1}^{N} I(X_i:\beta_i) \leq M.$ Both classical and quantum correlations have been proved to satisfy the IC condition \cite{nature}. It was further shown that, if Alice and Bob share arbitrary two inputs and two outputs no-signaling correlations corresponding to conditional probabilities $P_{ab|xy}$, then by applying a protocol by Van Dam \cite{vandam} and Wolf and Wullschleger \cite{wull}, one can derive a necessary condition for respecting the IC principle. The mathematical form of necessary conditions from Alice to Bob $(A \rightarrow B)$ can be expressed as:
 \begin{equation}\label{a-b}
   \sum_{y=0,1}\left[ \sum_{x=0,1}P(a\oplus b=xy\oplus \alpha x \oplus \beta y \oplus \gamma |xy)-1 \right]^2 \leq 1
 \end{equation}
and  the necessary conditions to satisfy IC from Bob to Alice $(B\rightarrow A)$ can be expressed as:
\begin{equation}\label{b-a}
  \sum_{x=0,1}\left[ \sum_{y=0,1}P(a\oplus b=xy\oplus \alpha x \oplus \beta y \oplus \gamma |xy)-1 \right]^2 \leq 1
 \end{equation}
where $ \alpha ,\beta ,\gamma \in \{0,1\} $. It is important to note  that the above conditions  are only a sufficient condition \cite{nature} for non-violating the IC principle. It means that a violation of ({\ref{a-b}}) or ({\ref{b-a}}) implies a violation of IC but the converse may not be true \cite{all09}. Notice that, the upper bound of success probability of the case HNA under imposing (NS+IC) is equal to be $0.2071$ which is same as that for the case of CNA \cite{ahanj} (please see the TABLE I).
\subsection{  Macroscopic Locality Principle }
 Migual Navascues and et.al \cite{navas07,navas08} introduced the macroscopic locality (ML) principle. The principle of ML states that two parties ( Alice and Bob) performing coarse-grained extensive measurements over independent correlated pairs of physical systems can always interpret their observations with a classical theory.  They proposed a way to approximate quantum correlations $Q$. They show that, there exist a series of correlation sets which, asymptotically converge on the set of quantum correlations. i.e. $Q$. The first step in this series of correlation sets, i.e. $Q_1$ exactly coincides the set correlations respecting ML principle and this set strictly contains the quantum set $Q$, thus showing  the insufficiency of ML principle in distinguishing all post-quantum correlations.\\Also, Navascues and et.al \cite{navas07,navas08} show that for a bipartite system with two binary inputs and two binary outputs exists a the necessary and sufficient criteria for respecting ML given by the condition:
\begin{equation}\label{ml}
|\sum_{x,y=0,1}(-1)^{xy}sin^{-1}\\(\frac{C_{xy}-C_x C_y}{\sqrt{(1-C_x^2)(1-C_y^2)}}\\)|\leq \pi
\end{equation}
where $C_{xy}=\sum_{a,b}(-1)^{a\oplus b} p_{ab|xy}$ ,$C_{x}=\sum_{a,b}(-1)^{a} p_{ab|xy}$ and $C_{y}=\sum_{a,b}(-1)^{ b} p_{ab|xy}$.\\
 S. Das and et.al \cite{lohardy} numerically showed that, the maximum of success probability HNA turn out to be $\approx 0.2062$ which is same as that for CNA under imposing $(NS+ML)$ condition (please see the TABLE I).
\subsection{Local Orthogonality Principle}
T. Fritz and et.al \cite{fritz} considered a pair of events $e=(a_1,...,a_n|x_1,...,x_n)$ and
 $ e\prime=(a\prime_1,...,a\prime_n|x\prime_1,...,x\prime_n)$ in a Bell scenario involving $n$ partite, with $m$ different inputs $x_i\in\{0,1,...,m-1\}$ and $d$ possible outputs $a_i\in\{0,1,...,d-1\}$ (where $~i\in \{0,1,...,n-1\}$). Next, they called that these two different events $e$ and $e\prime$ are  locally orthogonal, if  involve different outcomes of the same measurement by at least one party. The collection of events  $\{e_i\}$ are called orthogonal set, if they are  pairwise orthogonal and according to the LO principle,  the sum of probabilities of pairwise orthogonal event cannot exceed 1.
\begin{equation}\label{lo}
  \sum_{i} p(e_i)\leq 1
\end{equation}
Also, we can represent a set of local orthogonal events $\{e_i\}$ as a graph, where an event corresponds to a vertex and a pair of orthogonal vertex defines an edge and a set of orthogonal vertex forms a clique. So, any clique in the orthogonality graph of events is equivalent to an LO inequality (Eq \ref{lo}). On the other hand, it is worth that mention here, unlike IC and ML principles, there exist correlations such that single copy of those satisfy LO inequality, but two or more copies of those correlations violate LO principle. Also, the one copy of bipartite scenario, LO constraint is equivalent to NS conditions\cite{fritz}.\\More recently, S. Das and et.al \cite{lohardy} considered two orthogonal graph, that first graph contains $169$ vertices correspond to two copies of binary inputs-outputs Hardy correlation  and the second graph contains $196$ vertices corresponded to two copies of binary inputs-outputs Cabello correlation. Next, they showed that for getting the maximum success of HNA (CNA) under the full set of resulting LO inequalities, it is sufficient to maximize it under a small subset of LO inequalities contains 10 (13) inequalities.( please see the Appendices A and B ). Next, they \cite{lohardy}  obtained that by applying the NS principle and 10 LO inequalities, the maximum success probability in the case HNA  turns out to be 0.177 but it is not same as the bound of  CNA (please see the TABLE I).
\begin{table}
\label{tablenolr}
\caption{The Maximum value of HNA  and CNA under imposing the  NS, IC, ML and LO conditions. Our numerically calculation  by MATLAB software exactly confirm all the previous results.}
\centering
\begin{center}
    \begin{tabular}{| c |c | c | c | c | c | c|}

        \hline
   Case &$ LHVT$ &$ QM $ & $NS$ & $  NS+IC $ &$ NS+ML $ &  $ NS+LO $\\ \hline

    HNA &0&0.09  & 0.5 & 0.2071 & 0.206 & 0.177  \\ \hline \hline
    CNA &0&0.11  & 0.5 & 0.2071 & 0.206 & 0.2071  \\ \hline\hline

    \end{tabular}
\end{center}
\end{table}
\section{ The context of local randomness }
The local randomness condition imposes that the marginal probabilities of all possible outcomes on Alice's (Bob's) side for the $ x(y)$ input, are equal \cite{gazi}. So, in the case of two inputs and two outputs bipartite correlations, an input $x$ on Alice's side is locally random if for any choice of Bob's input $y$, we have:
\begin{eqnarray}
  P_{a=0|x}&=& \sum_b P_{0b|xy}=\frac{1}{2}~~(\mathrm{we~ show ~with ~0_A})\label{lr1}\\
  P_{a=1|x}&=& \sum_b P_{1b|xy}=\frac{1}{2}~~(\mathrm{we ~show~ with ~1_A})\label{lr2}.
\end{eqnarray}
Similarly, an input $y$ on Bob's side is locally random  if for any choice of Alice's input $x$, we have:
\begin{eqnarray}\label{random2}
  P_{b=0|y} &=& \sum_a P_{a0|xy}=\frac{1}{2}~\mathrm{(we~ show~ with~ 0_B)}\label{lr3} \\
  P_{b=1|y} &=& \sum_a P_{a1|xy}=\frac{1}{2}~ \mathrm{(we~ show~ with~ 1_B)}\label{lr4}.
\end{eqnarray}
In this stage, we want to find the maximum success probability of HNA and CNA under IC principle inequalities, ML principle inequality and under the full set of resulting LO inequalities, in the context of all the possibilities of local randomness.\\ Now, let us introduce the following maximization:\\
\textbf{Problem}\\
 Maximize $q_4=\frac{c_6}{2}$ ~( the case of HNA ) \\or\\ Maximize $q_4-q_1=\frac{c_6-c _{11}}{2}-c_{7,9,10}$ ~(the case of CNA ) \\
 Subject to the constraints:\\
 i) The positivity: Eq(\ref{positivity})\\
 ii) The normalization: Eq(\ref{normalization})\\
 iii) The NS conditions: Eqs(\ref{ns1},\ref{ns2})\\
 iv) The IC inequalities: Eqs (\ref{a-b},\ref{b-a})\\or\\ The ML inequality: Eq(\ref{ml})\\or\\
 The LO inequalities: Eqs (\ref{lohardy1}-\ref{lohardy2}) in the Appendix A for the  case of HNA
  \\or \\The LO inequalities: Eqs (\ref{locabelo1}-\ref{locabelo2}) in the Appendix B for the  case of CNA\\
 v) Without considering LR . \\
 vi) Under consideration LR: Eqs (\ref{lr1}-\ref{lr4})\\
 The optimal value of this problem gives us, an upper bound of the HNA and CNA for all possible choice of inputs that can be locally random. We solve this optimization problem  by using a program in MATLAB software (please see the appendix C). We present the our results for every choice of  collection $\{0_a, 1_A ,0_B, 1_B\}$ in the TABLES II, III.\\ Numerical calculation shows that the maximum value of CNA under the restrictions of IC, ML and LO is strictly larger than HNA under consideration LR. The distance between two random points $(x_1,...,x_n)$ and $(y_1,...,y_n)$ is defined by: $d(x,y)=\sqrt{\sum_{i=1}^{n}(x_i-y_i)^2}$, then we see that the distance of LO correlation from the QM correlations in both cases HNA  and CNA is less than IC and ML. On the other hand, the distance of IC, ML and LO correlations from the QM correlations in the case of CNA is much larger than HNA ( please see the TABLE IV).
 \begin{table*} [t]
\label{hnalr}
\caption{The numerical Maximum value of the HNA for the corresponding choice of inputs to be locally random and satisfy IC, ML and LO principles. }
%\centering

\begin{center}
    \begin{tabular}{|c |c| c | c | c | c | c |}

        \hline
   Case & Locally random inputs &$ Max (HNA)_{QM}$ & $Max (HNA)_{NS}$ & $  Max (HNA)_{NS+IC}$ &$ Max (HNA)_{NS+ML}$ &  $ Max (HNA)_{NS+LO}$\\ \hline

    1 &$0_A,1_A,0_B,1_B$  & 0 & 0.5 & 0 & 0 & 0 \\ \hline \hline
    2 &  $0_A,1_A,0_B $   & 0 & 0.5 &0.2071 &0.1002 & 0 \\ \hline
    3 &  $0_A,1_A,1_B $   & 0 & 0.5 &0.002 &0 & 0 \\ \hline
    4 &  $0_A,0_B,1_B $   & 0 & 0.5 &0.2071 &0.0998 & 0 \\ \hline
    5 &  $1_A,0_B,1_B $   & 0 & 0.5 &0.0016 &0 & 0 \\ \hline\hline
    6 &  $0_A,1_A $   & 0 & 0.5 &0.2071 &0.0967 & 0 \\ \hline
    7 &  $0_B,1_B $   & 0 & 0.5 &0.2071 &0.0967 & 0 \\ \hline
    8 &  $1_A,1_B $   & 0 & 0.5 &0 &0 & 0 \\ \hline
    9 &  $0_A,0_B $   & 0 & 0.5 &0.2071 &0.2 & 0 \\ \hline
    10 & $0_A,1_B $   & 0 & 0.5 &0.2071 &0.12 & 0 \\ \hline
    11&  $1_A,0_B $   & 0 & 0.5 &0.2071 &0.1045 & 0.1250 \\ \hline\hline
    12 &  $0_A $   & 0.0858 & 0.5 &0.2071 &0.2065 & 0.1760 \\ \hline
    13 &  $1_A $   & 0.0556 & 0.5 &0.2071 &0.1776 & 0.1344 \\ \hline
    14 &  $0_B $   & 0.0556 & 0.5 &0.2071 &0.2038 & 0.1692 \\ \hline
    15 &  $1_B $   & 0.0858 & 0.5 &0.2071 &0.1879 & 0.1508 \\ \hline\hline
    16 &  $NO ~LR $   & 0.09 & 0.5 &0.2071 &0.2063 & 0.1770 \\ \hline

    \end{tabular}
\end{center}

\end{table*}

\begin{table*}[t]
\label{cnalr}
\caption{The numerical Maximum value of the CNA for the corresponding choice of inputs to be locally random and satisfy IC, ML and LO principles.}
\centering

\begin{center}
    \begin{tabular}{| c |c | c | c | c | c |  c|}

        \hline
   Case & Locally random inputs &$ Max (CNA)_{QM}$ & $Max (CNA)_{NS}$ & $  Max (CNA)_{NS+IC}$ &$ Max (CNA)_{NS+ML}$ &  $ Max (CNA)_{NS+LO}$\\ \hline

    1 &$0_A,1_A,0_B,1_B$  & 0 & 0.5 & 0.2071 & 0.1020 & 0.1954 \\ \hline \hline
    2 &  $0_A,1_A,0_B $   & 0 & 0.5 &0.2071 &0.1940 & 0.1924 \\ \hline
    3 &  $0_A,1_A,1_B $   & 0 & 0.5 &0.2071 &0.1830 & 0.2 \\ \hline
    4 &  $0_A,0_B,1_B $   & 0 & 0.5 &0.2071 &0.1976 & 0.2071 \\ \hline
    5 &  $1_A,0_B,1_B $   & 0 & 0.5 &0.2071 &0.1845 & 0.2070 \\ \hline\hline
    6 &  $0_A,1_A $   & 0 & 0.5 &0.2071 &0.1963 & 0.2 \\ \hline
    7 &  $0_B,1_B $   & 0 & 0.5 &0.2071 &0.1976 & 0.2071 \\ \hline
    8 &  $1_A,1_B $   & 0 & 0.5 &0.2071 &0.1850 & 0.2071 \\ \hline
    9 &  $0_A,0_B $   & 0 & 0.5 &0.2071 &0.2034 & 0.2071 \\ \hline
    10 & $0_A,1_B $   & 0 & 0.5 &0.2071 &0.1976 & 0.1990 \\ \hline
    11&  $1_A,0_B $   & 0 & 0.5 &0.2071 &0.1949 & 0.1967 \\ \hline\hline
    12 &  $0_A $   & 0.0992 & 0.5 &0.2071 &0.2063 & 0.2071 \\ \hline
    13 &  $1_A $   & 0.0716 & 0.5 &0.2071 &0.1964 & 0.2071 \\ \hline
    14 &  $0_B $   & 0.0992 & 0.5 &0.2071 &0.2038 & 0.2071 \\ \hline
    15 &  $1_B $   & 0.0992 & 0.5 &0.2071 &0.1991 & 0.2071 \\ \hline\hline
    16 &  $NO ~LR $   & 0.1078 & 0.5 &0.2071 &0.2063 & 0.1978 \\ \hline

    \end{tabular}
\end{center}

\end{table*}

\begin{table*}[t]
\label{euclidean}
\caption{The Euclidean distance IC, ML and LO correlations from QM correlations under consideration LR }
\centering

\begin{center}
    \begin{tabular}{| c |c | c | c | c | }

        \hline
   Case &$ d(IC+NS,QM)$ &$d(ML+NS,QM) $ & $d(LO+NS,QM)$ \\ \hline

    HNA &0.6239  & 0.4233& 0.2377   \\ \hline \hline
    CNA &0.7362  & 0.6701 & 0.7183  \\ \hline\hline

    \end{tabular}
\end{center}
\end{table*}
\section{ Conclusion}
In this article, we study all the possibilities of local randomness in the HNA and the CNA respected by the principle of non-violation of IC, ML and LO inequalities. We obtain that without considering local randomness, the maximum success probability of the case HNA under imposing NS, IC, ML is equal to the case of CNA. Therefore, in this stage, there is no benefit between them. Next, we get interesting results, after considering all possibilities of the local randomness. We see that  the gap between QM and the above principles, in the context of the Cabello nonlocal argument is larger than the Hardy's case. This difference gap is interesting because may be relevant to assessing the viability of ``information causality" , ``macroscopic locality" and ``local orthogonality"  as partial candidate explanations for why QM correlations are weaker than generalized non-signalling correlations. So, in the case of CNA, the number of non-quantum correlation definitely obey the IC, ML, LO condition is more than the HNA case. Therefore,  we can conclude our work that the optimal success probability of CNA in QM is stronger control than the HNA for detecting post-quantum no signaling correlations. However, it remains to see, in the future, whether some stronger necessary condition  can  explain the upper bound of nonlocality in QM.
\section{Appendix A }
The following inequalities are sufficient for obtaining the upper bound of the HNA under LO principle\cite{lohardy}.
The inequalities can be shown, in terms of variables $e_k$, $m_i$ and $n_i$ ( $k\in\{1,2,3,4\}$ and $i\in\{1,2\}$).
\begin{eqnarray}
\label{lohardy1}
&&e_3^2+2e_2g_2-g_2^2-e_2^2\leq 0\\
&&e_3^2+2e_1g_2-e_1^2-g_2^2\leq 0\\
&&e_3^2+(e_3-e_1)(1-e_2-f_2)\leq 0\\
&&e_3^2+(e_3-e_2)(1-f_2-g_2) \leq 0\\
&&e_2(e_3+f_2-e_2)+(e_3-f_2)g_2\leq 0\\
&&e_1(f_2-e_3)+e_3(e_3+g_2)-f_2g_2\leq 0\\
&&(g_2-e_2)(f_2+g_2-1)+2e_1e_3-e_1^2\leq 0\\
&&e_3(1+f_2-g_2)+e_2(f_2+g_2-1)-f_2^2\leq  0\\
&&e_3^2+(f_2-e_3)(f_2+g_2-1)-(e_1-e_2)^2\leq 0\\
\label{lohardy2}
&&e_1(e_3-f_2-g_2)+e_2(-1+e_1-e_3+f_2+g_2)\nonumber\\&& +e_3\leq 0
\end{eqnarray}
  The relation between the 8 independent parameter  and the coefficients $c_i$ is obtained by comparing the matrix (\ref{matrix}) and the Hardy correlation matrix(\ref{hardy matrix}). So, we have:
  \begin{eqnarray*}
   e_1 &=& c_3+\frac{c_6}{2}\\
   e_2& =& c_{3,4}+\frac{c_6}{2} \\
   e_3& =& \frac{c_6}{2} \\
   e_4 &=& 0 \\
   f_1&=& e_2 \\
   f_2& =& c_5+\frac{c_{6}}{2} \\
   g_1 &=&e_1 \\
   g_2& =& c_2+e_2
  \end{eqnarray*}
\section{appendix B}
The following inequalities are sufficient for obtaining the upper bound of CNA under LO principle\cite{lohardy}. The inequalities can be shown, in terms of variables $e_k$, $m_i$ and $n_i$ ( $k\in\{1,2,3,4\}$ and $i\in\{1,2\}$).
\begin{eqnarray}
\label{locabelo1}
&&e_3(1+e_1-f_1-g_2)+(1+e_2+e_3-f_1-f_2-g_2)e_2\nonumber\\&&-e_1+(e_1+f_2+g_2-1)g_2\leq 0 \nonumber \\ \\
&&(e_2)^2+(1-f_1-g_2)e_3+(1+e_1+e_3-2f_1-f_2-g_2)e_2\nonumber\\&&-e_1^2+(e_3+f_1-f_2)e_1+(f_1+f_2)(f_2+g_2)-(f_1+f_2) \leq 0 \nonumber\\ \\
&&(e_2+e_3)(e_3-f_1)+2e_2g_2-g_2^2 \leq  0 \\
&&e_2^2+(e_2+2f_2-f_1)e_3-e_1 e_2+(1-f_2)e_1-\nonumber\\&&f_2^2+(f_2-1)f_1\leq 0 \nonumber\\ \\
&&e_2^2+e_3(1+f_2-f_1-g_2)+e_3e_2+e_2(1-f_1-f_2-g_2)\nonumber\\&&+e_1(g_2-f_1)+f_1^2+f_2(f_1+g_2-1)-f_1\leq 0\nonumber \\ \\
&&e_3^2+e_3(-f_1)+e_2e_3+e_2(2-2f_1-f_2)+\nonumber\\&&e_1(1-f_2-g_2)+f_1^2+f_1(2g_2+f_2-2)\nonumber\\&&+(f_2-1)g_2\leq  0 \\
&&e_3^2+(e_2-f_1)e_3+e_2(1-2f_1+g_2)+\nonumber\\&&f_1(f_1+g_2)-f_1-g_2^2\leq 0 \\
\label{locabelo2}
&&(e_2+e_1)(e_3+e_2+1-e_1-g_2)+f_1(f_1+2g_2-2e_2-2) \leq  0 \nonumber\\
\end{eqnarray}
The relation between the 8 independent parameters and the  coefficients $c_i$ is obtained by comparing the matrix (\ref{matrix}) and the Cabelo correlation matrix(\ref{cabelo matrix}). So, we have:
\begin{eqnarray*}
e_1&=&c_{3,8,10}+\frac{c_6+c_{11}}{2}\\
e_2&=&c_{3,4}+\frac{c_6}{2}\\
e_3&=&c_8+\frac{c_6}{2}\\
e_4&=&0\\
f_1&=&e_1+c_{4,7,9}\\
f_2&=&e_3+c_{5,7}+\frac{c_11}{2}\\
g_1&=&e_1\\
g_2&=&e_{2}+c_2+\frac{c_{11}}{2}
\end{eqnarray*}
\section{appendix c}
\subsection{The MATLAB program to find  the maximum violation HNA under IC, ML and LO under imposing LR}
fun=@(x)(-1)*(x(6)/2);\\
aeq0a1=[1 1 0 0 0 0.5];\\
aeq0a2=[0 0 1 1 1 0.5];\\
aeq0a=[aeq0a1;aeq0a2];\\
aeq1a1=[0 0 1 0 0 0.5];\\
aeq1a2=[1 1 0 1 1 0.5];\\
aeq1a=[aeq1a1; aeq1a2];\\
aeq0b1=[0 0 1 1 0 0.5];\\
aeq0b2=[1 1 0 0 1 0.5];\\
aeq0b=[aeq0b1;aeq0b2];\\
aeq1b1=[1 0 0 0 0 0.5];\\
aeq1b2=[0 1 1 1 1 0.5];\\
aeq1b=[aeq1b1; aeq1b2];\\
aeqcon=ones(1,6);\\
beqlr1=0.5*ones(8,1);\\
beqlr2=0.5*ones(6,1);\\
beqlr3=0.5*ones(4,1);\\
beqlr4=0.5*ones(2,1);\\
Aeq0=aeqcon;\\
beq0=1;\\
Aeq1=[aeqcon;aeq0a;aeq1a;aeq0b;aeq1b];\\
beq1=[1;beqlr1];\\
Aeq2=[aeqcon;aeq0a;aeq1a;aeq0b];\\
beq2=[1;beqlr2];\\
Aeq3=[aeqcon;aeq0a;aeq1a;aeq1b];\\
beq3=[1;beqlr2];\\
Aeq4=[aeqcon;aeq0a;aeq0b;aeq1b];\\
beq4=[1;beqlr2];\\
Aeq5=[aeqcon;aeq1a;aeq0b;aeq1b];\\
beq5=[1;beqlr2];\\
Aeq6=[aeqcon;aeq0a;aeq1a];\\
beq6=[1;beqlr3];\\
Aeq7=[aeqcon;aeq0b;aeq1b];\\
beq7=[1;beqlr3];\\
Aeq8=[aeqcon;aeq1a;aeq1b];\\
beq8=[1;beqlr3];\\
Aeq9=[aeqcon;aeq0a;aeq0b];\\
beq9=[1;beqlr3];\\
Aeq10=[aeqcon;aeq0a;aeq1b];\\
beq10=[1;beqlr3];\\
Aeq11=[aeqcon;aeq1a;aeq0b];\\
beq11=[1;beqlr3];\\
Aeq12=[aeqcon;aeq0a];\\
beq12=[1;beqlr4];\\
Aeq13=[aeqcon;aeq1a];\\
beq13=[1;beqlr4];\\
Aeq14=[aeqcon;aeq0b];\\
beq14=[1;beqlr4];\\
Aeq15=[aeqcon;aeq1b];\\
beq15=[1;beqlr4];\\
ub=ones(6,1);\\
lb=zeros(6,1);\\
%x0=(1/6)*[1 1 1 1 1 1];
x0=0.2*[1 1 1 1 1 0];
%x0=0.25*[1 1 1 1 0 0];
$ ml=@mlhardy;$\\
$ ic=@ichardy;$\\
$ lo=@lohardy;$\\
{fminlo0=fmincon(fun,x0,[],[],Aeq0,beq0,lb,ub,lo);\\
fminic0=fmincon(fun,x0,[],[],Aeq0,beq0,lb,ub,ic);\\
fminml0=fmincon(fun,x0,[],[],Aeq0,beq0,lb,ub,ml);\\
fminlo1=fmincon(fun,x0,[],[],Aeq1,beq1,lb,ub,lo);\\
fminic1=fmincon(fun,x0,[],[],Aeq1,beq1,lb,ub,ic);\\
fminml1=fmincon(fun,x0,[],[],Aeq1,beq1,lb,ub,ml);\\
fminlo2=fmincon(fun,x0,[],[],Aeq2,beq2,lb,ub,lo);\\
fminic2=fmincon(fun,x0,[],[],Aeq2,beq2,lb,ub,ic);\\
fminml2=fmincon(fun,x0,[],[],Aeq2,beq2,lb,ub,ml);\\
fminlo3=fmincon(fun,x0,[],[],Aeq3,beq3,lb,ub,lo);\\
fminic3=fmincon(fun,x0,[],[],Aeq3,beq3,lb,ub,ic);\\
fminml3=fmincon(fun,x0,[],[],Aeq3,beq3,lb,ub,ml);\\
fminlo4=fmincon(fun,x0,[],[],Aeq4,beq4,lb,ub,lo);\\
fminic4=fmincon(fun,x0,[],[],Aeq4,beq4,lb,ub,ic);\\
fminml4=fmincon(fun,x0,[],[],Aeq4,beq4,lb,ub,ml);\\
fminlo5=fmincon(fun,x0,[],[],Aeq5,beq5,lb,ub,lo);\\
fminic5=fmincon(fun,x0,[],[],Aeq5,beq5,lb,ub,ic);\\
fminml5=fmincon(fun,x0,[],[],Aeq5,beq5,lb,ub,ml);\\
fminlo6=fmincon(fun,x0,[],[],Aeq6,beq6,lb,ub,lo);\\
fminic6=fmincon(fun,x0,[],[],Aeq6,beq6,lb,ub,ic);\\
fminml6=fmincon(fun,x0,[],[],Aeq6,beq6,lb,ub,ml);\\
fminlo7=fmincon(fun,x0,[],[],Aeq7,beq7,lb,ub,lo);\\
fminic7=fmincon(fun,x0,[],[],Aeq7,beq7,lb,ub,ic);\\
fminml7=fmincon(fun,x0,[],[],Aeq7,beq7,lb,ub,ml);\\
fminlo8=fmincon(fun,x0,[],[],Aeq8,beq8,lb,ub,lo);\\
fminic8=fmincon(fun,x0,[],[],Aeq8,beq8,lb,ub,ic);\\
fminml8=fmincon(fun,x0,[],[],Aeq8,beq8,lb,ub,ml);\\
fminlo9=fmincon(fun,x0,[],[],Aeq9,beq9,lb,ub,lo);\\
fminic9=fmincon(fun,x0,[],[],Aeq9,beq9,lb,ub,ic);\\
fminml9=fmincon(fun,x0,[],[],Aeq9,beq9,lb,ub,ml);\\
fminlo10=fmincon(fun,x0,[],[],Aeq10,beq10,lb,ub,lo);\\
fminic10=fmincon(fun,x0,[],[],Aeq10,beq10,lb,ub,ic);\\
fminml10=fmincon(fun,x0,[],[],Aeq10,beq10,lb,ub,ml);\\
fminlo11=fmincon(fun,x0,[],[],Aeq11,beq11,lb,ub,lo);\\
fminic11=fmincon(fun,x0,[],[],Aeq11,beq11,lb,ub,ic);\\
fminml11=fmincon(fun,x0,[],[],Aeq11,beq11,lb,ub,ml);\\
fminlo12=fmincon(fun,x0,[],[],Aeq12,beq12,lb,ub,lo);\\
fminic12=fmincon(fun,x0,[],[],Aeq12,beq12,lb,ub,ic);\\
fminml12=fmincon(fun,x0,[],[],Aeq12,beq12,lb,ub,ml);\\
fminlo13=fmincon(fun,x0,[],[],Aeq13,beq13,lb,ub,lo);\\
fminic13=fmincon(fun,x0,[],[],Aeq13,beq13,lb,ub,ic);\\
fminml13=fmincon(fun,x0,[],[],Aeq13,beq13,lb,ub,ml);\\
fminlo14=fmincon(fun,x0,[],[],Aeq14,beq14,lb,ub,lo);\\
fminic14=fmincon(fun,x0,[],[],Aeq14,beq14,lb,ub,ic);\\
fminml14=fmincon(fun,x0,[],[],Aeq14,beq14,lb,ub,ml);\\
fminlo15=fmincon(fun,x0,[],[],Aeq15,beq15,lb,ub,lo);\\
fminic15=fmincon(fun,x0,[],[],Aeq15,beq15,lb,ub,ic);\\
fminml15=fmincon(fun,x0,[],[],Aeq15,beq15,lb,ub,ml);
[fminic0 fminic1 fminic2 fminic3 fminic4 fminic5 fminic6 fminic7; fminic8 fminic9,fminic10 fminic11 fminic12 fminic13 fminic14 fminic15;fminml0 fminml1 fminml2 fminml3 fminml4 fminml5 fminml6 fminml7;
fminml8 fminml9,fminml10 fminml11 fminml12 fminml13 fminml14 fminml15;
fminlo0 fminlo1 fminlo2 fminlo3 fminlo4 fminlo5 fminlo6 fminlo7;
 fminlo8 fminlo9,fminlo10 fminlo11 fminlo12 fminlo13 fminlo14 fminlo15]\\
 \subsubsection{@ichardy}
 function [c, ceq]=ichardy(x)\\
$p11=x(3)+x(6)/2;\\
p12=x(4);\\
p13=0;\\
p14=x(1)+x(2)+x(5)+x(6)/2;\\
p21=x(3)+x(4)+x(6)/2;\\
p22=0;\\
p23=x(2);\\
p24=x(1)+x(5)+x(6)/2;\\
p31=x(6)/2;\\
p32=x(5);\\
p33=x(3);\\
p34=x(1)+x(2)+x(4)+x(6)/2;\\
p41=0;\\
p42=x(5)+x(6)/2;\\
p43=x(3)+x(2)+x(4)+x(6)/2;\\
p44=x(1);\\
PI=(p11+p14+p31+p34)/2;\\
PII=(p21+p24+p42+p43)/2;\\
E1=(2*PI-1);\\
E2=(2*PII-1);\\
c1=E1^2+E2^2-1;\\
QI=(p11+p14+p21+p24)/2;\\
QII=(p31+p34+p42+p43)/2;\\
F1=(2*QI-1);\\
F2=(2*QII-1);\\
c2=F1^2+F2^2-1;\\
c=[c1;c2];$
ceq=[];
 \subsubsection{@mlhardy}
 function [c, ceq]=mlhardy(x)\\
p11=x(3)+x(6)/2;\\
p12=x(4);\\
p13=0;\\
$p14=x(1)+x(2)+x(5)+x(6)/2$;\\
$p21=x(3)+x(4)+x(6)/2$;\\
p22=0;\\
p23=x(2);\\
$p24=x(1)+x(5)+x(6)/2$;\\
p31=x(6)/2;\\
p32=x(5);\\
p33=x(3);\\
$p34=x(1)+x(2)+x(4)+x(6)/2$;\\
p41=0;\\
$p42=x(5)+x(6)/2$;\\
$p43=x(3)+x(2)+x(4)+x(6)/2$;\\
p44=x(1);\\
cx0=p11+p12-p13-p14;\\
cx1=p31+p32-p33-p34;\\
cy0=p11+p13-p12-p14;\\
cy1=p21+p23-p22-p24;\\
c00=p11+p14-p12-p13;\\
c01=p21+p24-p22-p23;\\
c10=p31+p34-p32-p33;\\
c11=p41+p44-p42-p43;\\
$d00=(c00-cx0*cy0)/(sqrt((1-(cx0)^2)*(1-(cy0)^2)))$;\\
$d01=(c01-cx0*cy1)/(sqrt((1-(cx0)^2)*(1-(cy1)^2)))$;\\
$d10=(c10-cx1*cy0)/(sqrt((1-(cx1)^2)*(1-(cy0)^2)))$;\\
$d11=(c11-cx1*cy1)/(sqrt((1-(cx1)^2)*(1-(cy1)^2)))$;\\
$c=abs(asin(d00)+asin(d01)+asin(d10)-asin(d11))-pi$;\\
ceq=[];
\subsubsection{@lohardy}
function [c, ceq]=lohardy(x)\\
p11=x(3)+x(6)/2;\\
p12=x(4);\\
p13=0;\\
%p14=x(1)+x(2)+x(5)+x(6)/2;
p21=x(3)+x(4)+x(6)/2;\\
%p22=0;\\
p23=x(2);\\
%p24=x(1)+x(5)+x(6)/2;
p31=x(6)/2;\\
p32=x(5);\\
%p33=x(3);\\
%p34=x(1)+x(2)+x(4)+x(6)/2;
p41=0;\\
%p42=x(5)+x(6)/2;
%p43=x(3)+x(2)+x(4)+x(6)/2;
%p44=x(1);
% matrix hardy has 8 independent parameters
%[c1 m0-c1 n0-c1 1+c1-m0-n0;
% c2 m0-c2 n1-c2 1+c2-m0-n1
% c3 m1-c3 n0-c3 1+c3-m1-n0
% c4 m1-c4 n1-c4 1+c4-m1-n1]
c1=p11;\\
c2=p21;\\
c3=p31;\\
%c4=p41;
%m0=p12+c1;
%n0=p13+c1;
n1=p23+c2;\\
m1=p32+c3;\\
$d1=c3^2+2*c1*n1-c1^2-n1^2$;\\
$d2=2*c1*c3-c1^2-(c2-n1)*(m1+n1-1)$;\\
$d3=c2*(c3+m1)+(c3-m1)*n1-c2^2$;\\
$d4=c3*(1+m1)+c2*(m1+n1)-c2-m1^2-c3*n1$;\\
$d5=c1*(m1-c3)+c3*(c3+n1)-m1*n1$;\\
$d6=c3+c1*c3+c2*(-1+c1-c3+m1+n1)-c1*(m1+n1)$;\\
$d7=c3+c3^2+c2*(m1+n1)-c2-c3*(m1+n1)$;\\
$d8=c3^2+2*c2*n1-c2^2-n1^2$;\\
$d9=c3*(1+c3)+c1*(c2+m1)-c1-c3*(c2+m1)$;\\
$d10=c3^2+m1*(-1+m1+n1)-(c1-c2)^2-c3*(-1+m1+n1)$;\\
c=[d1;d2;d3;d4;d5;d6;d7;d8;d9;d10];\\
ceq=[];
\subsection{The MATLAB program to find  the maximum violation CNA under IC, ML and LO under imposing LR}.
fun=@(x)(-1)*(0.5*(x(6)-x(11))-x(7)-x(9)-x(10));\\
%x0=(1/9)*[0 1 1 1 1 1 1 1 0 1 1];
%x0=[0 0.1 0.1 0.1 0.1 0.1 0.1 0.1 0.1 0.1 0.1];
%x0=(1/11)*[1 1 1 1 1 1 1 1 1 1 1];
x0=[1/8 1/8 1/8 1/8 1/8 1/8 1/20 1/20 1/20 1/20 1/20];\\
%A=-p22=-(x7+x8+x9+x10+x11/2)<-eps
A=(-1)*[0 0 0 0 0 0 1 1 1 1 0.5];\\
b=-eps;\\
aeqcon=ones(1,11);\\
beqcon=1;\\
%aeq0a1=x(1)+x(2)+x(7)+x(8)+x(9)+x(10)+0.5*(x(6)+x(11))=1/2;
%aeq0a2=x(3)+x(4)+x(5)+0.5*(x(6)+x(11));
aeq0a1=[1 1 0 0 0 0.5 1 1 1 1 0.5];\\
aeq0a2=[0 0 1 1 1 0.5 0 0 0 0 0.5];\\
aeq0a=[aeq0a1;aeq0a2];\\
%aeq1a1=x(1)+x(2)+x(4)+x(5)+x(7)+x(8)+0.5*(x(6)+x(11));
%aeq1a2=x(3)+x(9)+x(10)+0.5*(x(6)+x(11));
aeq1a1=[ 1 1 0 1 1 0.5 1 1 0 0 0.5];\\
aeq1a2=[ 0 0 1 0 0 0.5 0 0 1 1 0.5];\\
aeq1a=[aeq1a1;aeq1a2];\\
%aeq0b1=x(3)+x(4)+x(7)+x(8)+x(9)+x(10)+0.5*(x(6)+x(11));
%aeq0b2=x(1)+x(2)+x(5)+0.5*(x(6)+x(11));
aeq0b1=[ 0 0 1 1 0 0.5 1 1 1 1 0.5];\\
aeq0b2=[ 1 1 0 0 1 0.5 0 0 0 0 0.5];\\
aeq0b=[aeq0b1;aeq0b2];\\
%aeq1b1=x(2)+x(3)+x(4)+x(5)+x(7)+x(9)+0.5*(x(6)+x(11));
%aeq1b2=x(1)+x(8)+x(10)+0.5*(x(6)+x(11));
aeq1b1=[ 0 1 1 1 1 0.5 1 0 1 0 0.5];\\
aeq1b2=[ 1 0 0 0 0 0.5 0 1 0 1 0.5];\\
aeq1b=[aeq1b1;aeq1b2];\\
aeqlr=[aeq0a;aeq1a;aeq0b;aeq1b];\\
beqlr1=0.5*ones(8,1);\\
beqlr2=0.5*ones(6,1);\\
beqlr3=0.5*ones(4,1);\\
beqlr4=0.5*ones(2,1);\\
Aeq0=aeqcon;\\
beq0=beqcon;\\
Aeq1=[aeqcon;aeq0a;aeq1a;aeq0b;aeq1b];\\
beq1=[1;beqlr1];\\
Aeq2=[aeqcon;aeq0a;aeq1a;aeq0b];\\
beq2=[1;beqlr2];\\
Aeq3=[aeqcon;aeq0a;aeq1a;aeq1b];\\
beq3=[1;beqlr2];\\
Aeq4=[aeqcon;aeq0a;aeq0b;aeq1b];\\
beq4=[1;beqlr2];\\
Aeq5=[aeqcon;aeq1a;aeq0b;aeq1b];\\
beq5=[1;beqlr2];\\
Aeq6=[aeqcon;aeq0a;aeq1a];\\
beq6=[1;beqlr3];\\
Aeq7=[aeqcon;aeq0b;aeq1b];\\
beq7=[1;beqlr3];\\
Aeq8=[aeqcon;aeq1a;aeq1b];\\
beq8=[1;beqlr3];\\
Aeq9=[aeqcon;aeq0a;aeq0b];\\
beq9=[1;beqlr3];\\
Aeq10=[aeqcon;aeq0a;aeq1b];\\
beq10=[1;beqlr3];\\
Aeq11=[aeqcon;aeq1a;aeq0b];\\
beq11=[1;beqlr3];\\
Aeq12=[aeqcon;aeq0a];\\
beq12=[1;beqlr4];\\
Aeq13=[aeqcon;aeq1a];\\
beq13=[1;beqlr4];\\
Aeq14=[aeqcon;aeq0b];\\
beq14=[1;beqlr4];\\
Aeq15=[aeqcon;aeq1b];\\
beq15=[1;beqlr4];\\
ub=ones(11,1);\\
lb=zeros(11,1);\\
$ ic=@iccabelo;$\\
$ ml=@mlcabelo;$\\
$ lo=@locabelo;$\\
$
fminlo0=fmincon(fun,x0,[],[],Aeq0,beq0,lb,ub,lo);\\
fminic0=fmincon(fun,x0,[],[],Aeq0,beq0,lb,ub,ic);\\
fminml0=fmincon(fun,x0,[],[],Aeq0,beq0,lb,ub,ml);\\
fminlo1=fmincon(fun,x0,[],[],Aeq1,beq1,lb,ub,lo);\\
fminic1=fmincon(fun,x0,[],[],Aeq1,beq1,lb,ub,ic);\\
fminml1=fmincon(fun,x0,[],[],Aeq1,beq1,lb,ub,ml);\\
fminlo2=fmincon(fun,x0,[],[],Aeq2,beq2,lb,ub,lo);\\
fminic2=fmincon(fun,x0,[],[],Aeq2,beq2,lb,ub,ic);\\
fminml2=fmincon(fun,x0,[],[],Aeq2,beq2,lb,ub,ml);\\
fminlo3=fmincon(fun,x0,[],[],Aeq3,beq3,lb,ub,lo);\\
fminic3=fmincon(fun,x0,[],[],Aeq3,beq3,lb,ub,ic);\\
fminml3=fmincon(fun,x0,[],[],Aeq3,beq3,lb,ub,ml);\\
fminlo4=fmincon(fun,x0,[],[],Aeq4,beq4,lb,ub,lo);\\
fminic4=fmincon(fun,x0,[],[],Aeq4,beq4,lb,ub,ic);\\
fminml4=fmincon(fun,x0,[],[],Aeq4,beq4,lb,ub,ml);\\
fminlo5=fmincon(fun,x0,[],[],Aeq5,beq5,lb,ub,lo);\\
fminic5=fmincon(fun,x0,[],[],Aeq5,beq5,lb,ub,ic);\\
fminml5=fmincon(fun,x0,[],[],Aeq5,beq5,lb,ub,ml);\\
fminlo6=fmincon(fun,x0,[],[],Aeq6,beq6,lb,ub,lo);\\
fminic6=fmincon(fun,x0,[],[],Aeq6,beq6,lb,ub,ic);\\
fminml6=fmincon(fun,x0,[],[],Aeq6,beq6,lb,ub,ml);\\
fminlo7=fmincon(fun,x0,[],[],Aeq7,beq7,lb,ub,lo);\\
fminic7=fmincon(fun,x0,[],[],Aeq7,beq7,lb,ub,ic);\\
fminml7=fmincon(fun,x0,[],[],Aeq7,beq7,lb,ub,ml);\\
fminlo8=fmincon(fun,x0,[],[],Aeq8,beq8,lb,ub,lo);\\
fminic8=fmincon(fun,x0,[],[],Aeq8,beq8,lb,ub,ic);\\
fminml8=fmincon(fun,x0,[],[],Aeq8,beq8,lb,ub,ml);\\
fminlo9=fmincon(fun,x0,[],[],Aeq9,beq9,lb,ub,lo);\\
fminic9=fmincon(fun,x0,[],[],Aeq9,beq9,lb,ub,ic);\\
fminml9=fmincon(fun,x0,[],[],Aeq9,beq9,lb,ub,ml);\\
fminlo10=fmincon(fun,x0,[],[],Aeq10,beq10,lb,ub,lo);\\
fminic10=fmincon(fun,x0,[],[],Aeq10,beq10,lb,ub,ic);\\
fminml10=fmincon(fun,x0,[],[],Aeq10,beq10,lb,ub,ml);\\
fminlo11=fmincon(fun,x0,[],[],Aeq11,beq11,lb,ub,lo);\\
fminic11=fmincon(fun,x0,[],[],Aeq11,beq11,lb,ub,ic);\\
fminml11=fmincon(fun,x0,[],[],Aeq11,beq11,lb,ub,ml);\\
fminlo12=fmincon(fun,x0,[],[],Aeq12,beq12,lb,ub,lo);\\
fminic12=fmincon(fun,x0,[],[],Aeq12,beq12,lb,ub,ic);\\
fminml12=fmincon(fun,x0,[],[],Aeq12,beq12,lb,ub,ml);\\
fminlo13=fmincon(fun,x0,[],[],Aeq13,beq13,lb,ub,lo);\\
fminic13=fmincon(fun,x0,[],[],Aeq13,beq13,lb,ub,ic);\\
fminml13=fmincon(fun,x0,[],[],Aeq13,beq13,lb,ub,ml);\\
fminlo14=fmincon(fun,x0,[],[],Aeq14,beq14,lb,ub,lo);\\
fminic14=fmincon(fun,x0,[],[],Aeq14,beq14,lb,ub,ic);\\
fminml14=fmincon(fun,x0,[],[],Aeq14,beq14,lb,ub,ml);\\
fminlo15=fmincon(fun,x0,[],[],Aeq15,beq15,lb,ub,lo);\\
fminic15=fmincon(fun,x0,[],[],Aeq15,beq15,lb,ub,ic);\\
fminml15=fmincon(fun,x0,[],[],Aeq15,beq15,lb,ub,ml);$
-[fminic0 fminic1 fminic2 fminic3 fminic4 fminic5 fminic6 fminic7;
 fminic8 fminic9,fminic10 fminic11 fminic12 fminic13 fminic14 fminic15;
fminml0 fminml1 fminml2 fminml3 fminml4 fminml5 fminml6 fminml7;
fminml8 fminml9,fminml10 fminml11 fminml12 fminml13 fminml14 fminml15;
fminlo0 fminlo1 fminlo2 fminlo3 fminlo4 fminlo5 fminlo6 fminlo7;
 fminlo8 fminlo9,fminlo10 fminlo11 fminlo12 fminlo13 fminlo14 fminlo15]
\subsubsection{ @iccabelo};
function [c, ceq]=iccabelo(x)\\
p11=x(3)+x(8)+x(10)+(x(11)+x(6))/2;\\
p12=x(4)+x(7)+x(9);\\
p13=0;\\
p14=x(1)+x(2)+x(5)+(x(11)+x(6))/2;\\
p21=x(3)+x(4)+x(6)/2;\\
p22=x(7)+x(8)+x(9)+x(10)+x(11)/2;\\
p23=x(2)+x(11)/2;\\
p24=x(1)+x(5)+x(6)/2;\\
p31=x(8)+x(6)/2;\\
p32=x(5)+x(7)+x(11)/2;\\
p33=x(3)+x(10)+x(11)/2;\\
p34=x(1)+x(2)+x(4)+x(9)+x(6)/2;\\
p41=0;\\
p42=x(5)+x(7)+x(8)+(x(6)+x(11))/2;\\
p43=x(3)+x(2)+x(4)+(x(6)+x(11))/2;\\
p44=x(1)+x(9)+x(10);\\
EI=(p11+p14+p31+p34-1);\\
EII=(p21+p24+p42+p43-1);\\
$c1=EI^2+EII^2-1$;\\
$EIII=(p11+p14+p21+p24-1)$;\\
EIV=(p31+p34+p42+p43-1);\\
$c2=EIII^2+EIV^2-1$;\\
FI=(p11+p14+p32+p33-1);\\
FII=(p22+p23+p42+p43-1);\\
$c3=FI^2+FII^2-1;$\\
FIII=(p11+p14+p22+p23-1);\\
FIV=(p32+p33+p42+p43-1);\\
$c4=FIII^2+FIV^2-1$;\\
c=[c1;c2;c3;c4];\\
ceq=[];
\subsubsection{@mlcabelo};
function [c, ceq]=mlcabelo(x)\\
p11=x(3)+x(8)+x(10)+(x(11)+x(6))/2;\\
p12=x(4)+x(7)+x(9);\\
p13=0;\\
p14=x(1)+x(2)+x(5)+(x(11)+x(6))/2;\\
p21=x(3)+x(4)+x(6)/2;\\
p22=x(7)+x(8)+x(9)+x(10)+x(11)/2;\\
p23=x(2)+x(11)/2;\\
p24=x(1)+x(5)+x(6)/2;\\
p31=x(8)+x(6)/2;\\
p32=x(5)+x(7)+x(11)/2;\\
p33=x(3)+x(10)+x(11)/2;\\
p34=x(1)+x(2)+x(4)+x(9)+x(6)/2;\\
p41=0;\\
p42=x(5)+x(7)+x(8)+(x(6)+x(11))/2;\\
p43=x(3)+x(2)+x(4)+(x(6)+x(11))/2;\\
p44=x(1)+x(9)+x(10);\\
cx0=p11+p12-p13-p14;\\
cx1=p31+p32-p33-p34;\\
cy0=p11+p13-p12-p14;\\
cy1=p21+p23-p22-p24;\\
c00=p11+p14-p12-p13;\\
c01=p21+p24-p22-p23;\\
c10=p31+p34-p32-p33;\\
c11=p41+p44-p42-p43;\\
$d00=(c00-cx0*cy0)/(sqrt((1-(cx0)^2)*(1-(cy0)^2)));$\\
$d01=(c01-cx0*cy1)/(sqrt((1-(cx0)^2)*(1-(cy1)^2)));$\\
$d10=(c10-cx1*cy0)/(sqrt((1-(cx1)^2)*(1-(cy0)^2)));$\\
$d11=(c11-cx1*cy1)/(sqrt((1-(cx1)^2)*(1-(cy1)^2)));$\\
$c=abs(asin(d00)+asin(d01)+asin(d10)-asin(d11))-pi;$\\
ceq=[];
\subsubsection{@locabelo};
function [c,ceq]=locabelo(x)\\
e1=x(3)+x(8)+x(10)+(x(11)+x(6))/2;\\
e2=x(3)+x(4)+x(6)/2;\\
e3=x(8)+x(6)/2;\\
%e4=0;
f1=e1+x(4)+x(7)+x(9);\\
f2=e3+x(5)+x(7)+x(11)/2;\\
%g1=e1;
g2=x(2)+x(3)+x(4)+(x(6)+x(11))/2;\\
$ d1=e3*(1+e1-f1-g2)+(1+e2+e3-f1-f2-g2)*e2-e1+(e1+f2+g2-1)*g2;$\\
$d2=e2^2+(1-f1-g2)*e3+(1+e1+e3-2*f1-f2-g2)*e2-e1^2+(e3+f1-f2)*e1+(f1+f2)*(f2+g2-1);$\\
$d3=(e2+e3)*(e3-f1)+2*e2*g2-g2^2;$\\
$d4=e2^2+(e2+2*f2-f1)*e3-e1*e2+(1-f2)*(e1-f1)-f2^2;$\\
$d5=e2^2+e3*(1+f2-f1-g2)+e3*e2+e2*(1-f1-f2-g2)+e1*(g2-f1)+f1^2+f2*(f1+g2-1)-f1;$\\
$d6=e3^2+e3*(e2-f1)+e2*(2-2*f1-f2)+e1*(1-f2-g2)+f1^2+f1*(2*g2+f2-2)+(f2-1)*g2;$\\
$d7=e3^2+(e2-f1)*e3+e2*(1-2*f1+g2)+f1*(f1+g2-1)-g2^2;$\\
$d8=(e2+e1)*(e3+e2+1-e1-g2)+f1*(f1+2*g2-2*e2-2);$\\
$c=[d1;d2;d3;d4;d5;d6;d7;d8];$\\
ceq=[];
%%%%%%%%%%%%%%%%%%%%%%%%%%%%%%%

\end{document}